# Simple arithmetic operation in latent space can generate a novel three-dimensional graph metamaterials

*Namjung Kim[1], Dongseok Lee[2], Chanyoung Kim[2], Dosung Lee[1] and *Youngjoon Hong[2]

[1]Department of Mechanical Engineering, Gachon University, Sungnam, 13120, Korea

[2]Department of Mathematics, Korea Advanced Institute of Science and Technology, Daejeon, 34141, Korea

*Corresponding author
Prof. Youngjoon Hong:   hongyj@kaist.ac.kr
Prof. Namjung Kim:   namjungk@gachon.ac.kr

# Abstract

Recent advancements in artificial intelligence (AI)-based design strategies for metamaterials have revolutionized the creation of customizable architectures spanning nano- to macro-scale dimensions, achieving unprecedented mechanical behaviors that surpass the inherent properties of the constituent materials. However, the growing complexity of these methods poses challenges in generating diverse metamaterials without substantial human and computational resources, hindering widespread adoption. Addressing this, our study introduces an innovative design strategy capable of generating various three-dimensional graph metamaterials using simple arithmetic operations within the latent space. By seamlessly integrating hidden representations of disentangled latent space and latent diffusion processes, our approach provides a comprehensive understanding of complex design spaces, generating diverse graph metamaterials through arithmetic operations. This methodology stands as a versatile tool for creating structures ranging from repetitive lattice structures to functionally graded mechanical metamaterials. It also serves as an inverse design strategy for diverse lattice structures, including crystalline structures and those made of trabecular bone. We believe that this methodology represents a foundational step in advancing our comprehension of the intricate latent design space, offering the potential to establish a unified model for various traditional generative models in the realm of mechanical metamaterials.

# Introduction

In recent years, the field of metamaterials has undergone a transformative shift, marked by an escalating interest in developing materials with unprecedented properties and functionalities, including negative compressibility[1,2], negative Poisson's ratio[3–5], negative thermal expansion[6,7], ultra-lightness, ultra-stiffness, shape recoverability[8,9], high energy absorption rates[10,11], and programmable responses[12,13]. Such advancements have propelled metamaterials to the forefront of various high-impact applications, spanning high-performance sensing[14,15], energy harvesting[16], highly efficient actuation[17,18], and the storage of digital information[19,20]. Recently, applications of metamaterials have expanded to fields requiring extraordinary characteristics of the static and dynamic responses in mechanics[21,22], optics[23], and electromagnetics[24], positioning them as next-generation materials. This surge in metamaterials research reflects a broader trend in materials science, where the boundaries of possibility are continually being expanded through innovative approaches. Moreover, the unique properties of these materials have opened up new avenues in fields as diverse as medical technology, aerospace engineering, and sustainable energy solutions, underscoring their potential to address some of the most pressing challenges of our time.

The increasing demand for metamaterials with tailored and unprecedented behaviors underscores the imperative for a robust and comprehensive design methodology. This need is particularly pronounced in the realm of mechanical metamaterials, where traditional design paradigms — spanning heuristic, optimization-based[25–27], and nature-inspired approaches[28,29] — have been foundational yet remain insufficient if seeking to exploit the burgeoning potential of these materials fully. In this context, methodologies based on artificial intelligence (AI) have emerged as groundbreaking tool, revolutionizing the design process of mechanical metamaterials. AI methodologies have not only expedited the design process but have also enhanced the precision and scope of these designs, enabling both forward and inverse design strategies[30]. This paradigm shift is underpinned by AI's ability to navigate and interpret the vast design space of metamaterials, a task that has traditionally been challenging due to the inherent complexity[31–34] and multidimensional nature of metamaterials[35–38]. Recent advancements in AI-based design methodology range from forward and inverse designs of metamaterials[39–41] to a broader understanding of the constitutive relationships among base materials[42–44], covering the entire set of design components of mechanical metamaterials.

In AI-based design strategies, utilizing latent space stands out as a prominent method for generating innovative mechanical metamaterials. In the context of AI, the latent space is a lower-dimensional space where intricate input data is condensed into a concise and meaningful representation. The utilization of latent space offers several advantages, including dimensionality reduction, enhancement in design flexibility, and effective feature representation. These advantages not only streamline the design process but also deepen our understanding of the physical properties inherent in metamaterials. As the complexity of target systems increases, the significance of effective feature representation and dimensional reduction provided by latent space becomes increasingly evident. Recent research endeavors have introduced diverse frameworks aimed at harnessing the benefits of latent space in the design of mechanical metamaterials. Some studies focus on interpreting the physical implications of the latent space for metamaterial design[38,45], while others aim to unify its complexity to model both linear and nonlinear behaviors[31]. Additionally, efforts are directed towards inverse designing of nonlinear mechanical metamaterials[33], devising size-agnostic models for inverse design tasks[46], and dynamically modulating metamaterial responses[47]. However, challenges persist, including the latent space's intricate nature leading to a lack of interpretability, limitations in parametric representation for geometry, and an overreliance on extensive databases. Addressing these hurdles is pivotal for propelling the continued advancements of deep generative models in the field of mechanical metamaterials.

In this study, we propose a novel and data-efficient design framework capable of generating three-dimensional graph metamaterials using simple arithmetic operations within the latent space. Our model integrates two cutting-edge deep generative models with hidden representations of a disentangled latent space, ensuring both prediction accuracy and generalizability. The most important are summarized below:

1. Utilizing simple arithmetic operations in latent space, various graph metamaterials can be generated both within and beyond the training dataset, significantly reducing the reliance on human and computational resources.
2. The efficiency of the proposed framework surpasses that of previous deep generative models in terms of data utilization, achieving a reduction in training costs ranging from 20 to 200 fold.
3. By seamlessly integrating two state-of-the-art deep generative models, our framework facilitates the creation of diverse graph metamaterials with varying mechanical properties and enables the inverse design of graph metamaterials with specific target properties.
4. In comparison to other AI algorithms, the developed regressor demonstrates superior accuracy in predicting the mechanical properties of unseen graph metamaterials.
5. Through experimental validation, the feasibility of the generated structures is confirmed, showcasing the ability to achieve tailored mechanical behaviors using additive manufacturing technology.

Consequently, our model facilitates the rapid creation of metamaterials with fully customizable mechanical behaviors via simple arithmetic operations, achieving a harmonious equilibrium between the simplicity and complexity inherent in deep generative models. We believe that this breakthrough in understanding the latent space and its manipulation opens new horizons in the field. It paves the way for sophisticated manipulation and generation of complex and previously unseen graph structures with simple arithmetic operations, thereby significantly broadening the scope for both exploitation and exploration of desired graph configurations.

**Overview of the Framework for the Deep Generative Model**

Figure 1 (a) displays a foundational geometry, a basis unit cell, for the database. 5×5×5 grey nodes in a dimensionless design space serve as control nodes, and connecting edges between these nodes define the unit cell's geometry. This configuration, balancing expressiveness and computational efficiency, enables the creation of a wide range of metamaterial designs, from simple cubic structures to more intricate geometries such as tetrahedron-based or truncated cubic cells. Each unit cell comprises a node set $\boldsymbol{V} = \{\boldsymbol{v_1}, \boldsymbol{v_2}, \cdots, \boldsymbol{v_N}\}$, and an adjacency matrix $A$, which is a square N-dimensional matrix, consisting of 0s and 1s that represents undirected graphs. Additionally, each node in $V$ possesses corresponding features $X$, capturing the unique characteristics of each node such as positions. A family of 14 strut-based unit cells with cubic symmetry is chosen as basis cells (Figure S1 in Supporting information). This facilitates the generation of 16,383 structures through linear combinations as shown in Figure 1 (b). Despite the inherent limitations of cubic symmetry, our model significantly broadens the spectrum of achievable asymmetric properties, aided by a novel latent perturbation algorithm. Notably, the used database is highly compact in the order of 20 to 200, comparing to recent works on 3D unit cell generation[31,48,49]. However, our model's sophisticated balancing strategy allows for the precise generation of viable design candidates without the need for a substantial database.

The subsequent phase involves training a discretized latent space to capture representative features of the mechanical metamaterials. For each unit cell, the latent vector is set as N×18 vector, optimizing topological and geometrical expressiveness while minimizing computational demand. The proposed methodology addresses the bias-variance tradeoff through two distinct strategies: continuous interpolation to address bias, and generative denoising for variance mitigation. The continuous interpolation, utilizing a discrete latent representation embedded within the framework of a Variational Autoencoder (VAE) and the graph neural networks, yields transitional structures with stable topologies. Concurrently, the generative denoising strategy employs a latent diffusion model, integrating deterministic reverse diffusion processes alongside graph neural networks. This strategy transitions from a low likelihood initial design to a target feature, mitigating instability introduced by Gaussian random latent vectors through diffusion guidance.

This dual-phase methodology synergistically merges discrete latent representation with a score-based latent diffusion framework, harnessing the strengths of both VAE architecture and latent diffusion models. Our strategy ensures the accurate generation of viable designs, underscoring its efficiency relative to existing methodologies in 3D unit cell generation. Figure 1 (c) demonstrates the overview of the proposed framework and the comprehensive details, and its training strategies are elaborated in Supporting Information. Rigorous comparative analysis between VAE and latent diffusion model in Supporting Information reveals that our dual phase methodology is inevitable to balance the bias-variance tradeoff.

**Generation via Simple Arithmetic Operations in Disentangled Latent Space**

Minimizing the challenges associated with intricate manipulation in the latent space not only optimizes the design workflow but also fosters greater adaptability, enabling the systematic creation of materials with precise mechanical characteristics. Figure 2 elucidates the synthetic property of the latent space, employing a bubble plot to represent the discrete latent vectors and their associated graph structures. Let us assume $G_1$ and $G_2$ are two distinct graphs, $B_1$ and $B_2$ are corresponding latent basis, and $P_1$ and $P_2$ are corresponding positional matrix, where $B_1, B_2 \in \mathbb{R}^{N \times 18}$ and $P_1, P_2 \in \mathbb{R}^{N \times 3}$. The fundamental arithmetic principle is valid when the nodes of the two graphs do not overlap: $B_1 + B_2 = G_1 \oplus G_2$, where operation $\oplus$ stands for a union of two graphs. In the event of overlapping nodes between the two graphs, the principle persists flawlessly: if the overlapped set of position rows consists of $\{p_{1_1}, p_{1_2}, \dots p_{1_n}\}$ for $P_1$ and $\{p_{2_1}, p_{2_2}, \dots p_{2_n}\}$ for $P_2$, then the interpolated set between $\{b_{1_1}, b_{1_2}, \dots b_{1_n}\}$ for $B_1$ and $\{b_{2_1}, b_{2_2}, \dots b_{2_n}\}$ for $B_2$ becomes $\{\bar{b}_1, \bar{b}_2, \dots \bar{b}_n\}$, where $\bar{b}_k = (1-\alpha)b_{1_k} + \alpha b_{2_k}, 0 \leq \alpha \leq 1$. The fundamental arithmetic principle is still valid even if the overlapped latent vectors of $B_1$ and $B_2$ are changed by interpolation, denoting $\bar{B}_1$ and $\bar{B}_2$, then $\bar{B}_1 + \bar{B}_2 = G_1 \oplus G_2$. This latent operation also holds for the representation of triple graphs and basis, $B_1, B_2$ and $B_3$ such that $B_1 + B_2 + B_3 = G_1 \oplus G_2 \oplus G_3$.

Experimental evidence confirming the validity of simple arithmetic operations within the latent space is presented in Figure 4. The results demonstrate the principle's applicability in both overlapped (①) and non-overlapped (②) scenarios. Furthermore, comprehensive validation is performed across diverse combinations of two graphs (③ to ⑥). Intriguingly, our discoveries extend to representations involving not only two but three or more latent vectors. As demonstrated in the triple combination depicted in Figure 2, despite varying locations of overlapping nodes, the concatenation of latent vectors across all three graphs precisely yields the compound structure found in our database. Through experimental exploration of various combinations of basis unit cells, we deciphered that the simple arithmetic operations in latent space can generate various graph structures. This deep understanding of the latent space reveals that each basis unit cell correlates with a distinct latent space representation. Notably, when arithmetic operation is applied to the latent vectors of each basis graph, the resulting samples accurately reflect the combinations of the basis graph. To the best of our knowledge, this represents the inaugural demonstration of a straightforward operation capable of generating graph-based metamaterials.

**Physical Interpretation of Forward and Reverse Diffusion Process**

Although simple arithmetic operations can generate diverse metamaterials, the imperative to produce metamaterials with specific properties remains significant. To address this need, we employed a diffusion process with guidance in latent space to facilitate the design of metamaterials with targeted properties. During this process, the physical interpretation of the diffusion process is critical to both forward and inverse design capabilities. In the context of forward design, a sound physical interpretation ensures alignment with real-world mechanical principles, while in inverse design, it facilitates the precise specification and understanding of desired mechanical properties. By embedding physical insights into the model, the diffusion process becomes a reliable tool for systematically exploring the design space, producing structures that not only exhibit diverse properties but are also grounded in the principles of mechanics. Therefore, further investigation is performed to elucidate the diffusion process for generating unseen mechanical metamaterials, as shown in Figure 3.

In Figure 3, we present the distributions of $C_{11}$ values — original (blue), noised (grey), and reconstructed (green) datasets. The forward diffusion, marked by the transition from blue to grey, involves systematically dismantling the metamaterials' graph structure, eventually reaching to a noised state. This is evident in the discretized latent vectors' distribution, initially well-structured and localized, becoming increasingly dispersed throughout the process. The forward diffusion introduces white noise, defined as $z_t = \sqrt{\alpha_t} z_0 + \sqrt{1 - \alpha_t} \epsilon$ where $\epsilon \sim \mathcal{N}(0, I)$, and $\alpha_t$ is a noise weighting factor. $v_0$ and $v_t$ are the graph structure at the initial and the step t, respectively. Each noised latent vector is sampled from the probability distribution, $p(z_t | z_0) = \mathcal{N}(\sqrt{\alpha_t} z_0, (1 - \alpha_t) I)$, corresponding to the dismantled graph structure. This culminates in the convergence of $C_{11}$ values to a normal distribution.

The reverse diffusion process involves a mapping function from $z_t$ to $z_0$ to reconstruct the metamaterials' design. The mapping function is defined as $z_{t-1} = \sqrt{\alpha_{t-1}} \left( \frac{z_t - \sqrt{1-\alpha_t} \epsilon_\theta(z_t)}{\sqrt{\alpha_t}} \right) + \sqrt{1 - \alpha_{t-1}} \epsilon_\theta(z_t)$. The green distribution reflects the $C_{11}$ values of reconstructed metamaterials, obtained using a numerical solver[50]. Despite some minor discrepancies between the original and reconstructed distributions owing to a limited number of sampling points, the process effectively retains the unique characteristics of the metamaterials. The gradation from grey to red in the arrow highlights the transformation from noise level to the generation of novel designs with specific mechanical properties. Unseen designs emerge from the latent vector $v_t$, sampled from $\mathcal{N}(0,I)$, with the discretized latent vectors converging to targeted properties of 0.28 and 0.66, indicating the feasibility of the resulting graph structures (additional details are provided in Supporting Information). This convergence exemplifies the proposed model's prowess in inverse design, capable of generating novel designs from a noised state. The versatility of the model not only demonstrates its potential in scientific and engineering realms but also signals a paradigm shift in the approach to material design and optimization. This model stands as a cornerstone in accelerating the engineering discovery pipeline, illustrating our commitment to advancing the frontiers of material science. The details of geometrical evolution in noising and denoising processes are illustrated in Supporting Information.

**Evaluating Regression Accuracy**

To assess the regressor's efficacy within the proposed model for predicting the elasticity of metamaterials, we conducted a regression accuracy evaluation using graphs subjected to noise perturbation. Specifically, we illustrated the generation of a noise-affected graph sample, denoted as $G_{t-1}$, originating from a distribution $z_T \sim \mathcal{N}(0, \mathrm{I})$, through an iterative process given by $z_{t-1} = \sqrt{\alpha_{t-1}} f_\theta(z_t) + \sqrt{1 - \alpha_{t-1}} \epsilon_\theta(z_t)$, where $G_{t-1} = dec(z_{t-1})$, $dec(\cdot)$ is decoder operator, and $f_\theta(z_t) = \frac{z_t - \sqrt{1-\alpha_t} \epsilon_\theta(z_t)}{\sqrt{\alpha_t}}$. This iterative process demonstrates the regressor's proficiency at effectively managing and interpreting graph-based metamaterials subject to noise. Additional details are explained in Supporting Information. As depicted in Figure 4 (a), our regression analysis focused on noise-free graphs, utilizing a dataset comprising $G_0$ and the corresponding $C_{11}$ calculated via a numerical solver. The analysis highlighted the exceptional proficiency of our proposed regressor, surpassing traditional analytical models such as support vector machines (SVMs), Gaussian process regression (GPR), and ensemble methods in predictive accuracy. While conventional models tended to deviate from the ideal $y = x$ line, the predictions of the graph Transformer exhibited remarkable alignment with this line, indicative of its superior predictive precision compared to established methods. In Figure 4 (b), our analysis examined datasets with noisy graphs over the interval 0 < t < 100 during the diffusion process. Regression analysis revealed that the proposed regressor maintained a superior alignment with the $y = x$ line, signifying significantly enhanced predictive capability. Conversely, traditional models exhibited notable deviations in predictive accuracy as the noise level increased. These findings validate the robustness of our proposed regressor in effectively mitigating the disruptions caused by noise.

**Exploration and Exploitation in Property Space**

In the realm of mechanical metamaterials, exploration and exploitation in property space are paramount for harnessing their full spectrum of capabilities. Exploration allows designers to venture into unprecedented areas beyond the existing property space, fostering innovation and the creation of novel metamaterial structures. Conversely, exploitation involves refining designs based on existing knowledge and successful outcomes, tailoring designs to achieve precise mechanical properties for specific applications. This balance not only furthers design refinement but also optimizes metamaterials for targeted functionalities. Figure 5 (a) illustrates the expanded property space achieved through our model. The generation process consists of two distinct strategies: discrete latent space perturbation and guidance generation. As depicted in the figure, the distribution of directional elastic moduli of the metamaterials, normalized by the elastic modulus of the base material ($E_s$), spans a significantly broader range. This confirms that out model effectively generates the anisotropic structures from the isotropic database, expanding the feasible properties of metamaterials. Figure 5 (b) further illustrates our model's adeptness at controlling anisotropy. Through varying degrees of latent space perturbation, displayed in five distinct levels (Case A to E), we highlight the model's proficiency in generating asymmetric metamaterials from a database primarily composed of symmetric designs. The elastic surface, projected on the $C_{11}$-$C_{22}$ plane, exhibits an elastic modulus ratio varying from 1 to 2.75, offering quantitative guidance for achieving feasible asymmetry in the metamaterials. The details are explained in Supporting Information. While isotropic mechanical metamaterial designs are suitable for conventional applications[51], our model's capacity to tailor strong anisotropic properties opens new avenues for applications necessitating directional load endurance or materials exhibiting multifunctional and multistable behaviors[52].

Figure 5 (c) delineates the $C_{11}$ value distribution of our generated metamaterials overlaid upon the original database. The distinct and concentrated distribution in the elastic moduli of the metamaterials, marked in red and blue for normalized targeted $C_{11}$ values of 0.10 and 0.24 respectively, attests to the precision of our guidance generation. These newly generated mechanical metamaterials fill the gap in the training dataset, introducing novel designs absent in the original database. Various designs showcase varied structural forms tailored to specific elastic properties, diverging significantly from database counterparts with analogous properties (additional details are provided in Supporting Information). This diversity proves our model's ability in exploiting to create multiple metamaterials with distinct densities and geometries yet identical elastic properties. Our model's unique ability to manipulate elastic modulus independent of structural density disrupts the conventional correlation between these two attributes. This breakthrough offers unparalleled flexibility in metamaterial design, carving a unique niche in multi-objective and multifunctional application domains.

**Inverse Design for Various Lattice Structures**

Figure 6 highlights the extensive expressive capability of our model in addressing the intricacies of inverse design in mechanical metamaterials. This design paradigm, which meticulously tailors structures to align with specific performance criteria, has been revolutionized by the advent of deep generative models. Characterized by their proficiency in deciphering intricate patterns from extensive datasets, these models offer a unique avenue to transcend traditional design limitations. Our approach harmoniously integrates latent diffusion with guidance and discrete latent vector perturbation, adeptly navigating the design landscape of both isotropic and anisotropic materials, extending well beyond the confines of our existing database.

The figure displays exemplary unit cells with a wide range of elasticity and iso/anisotropic characteristics, featuring four representative examples of trabecular bones and crystallographic periodic networks from references[53,54](additional examples are included in Supporting Information). Graphs generated by diffusion guidance and those obtained through latent perturbation are colored in blue and pink, respectively. Relative elasticity surfaces, normalized by the elastic modulus of their base material, along with their projections onto $e_1$-$e_2$, $e_1$-$e_3$, and $e_2$-$e_3$ planes, are presented. Each axis ($e_1$, $e_2$, $e_3$) is normalized by the Young's modulus ($E_s$) of the corresponding topology. The projections of the reference material and the generated material are represented by the red and blue curves, respectively. Additionally, we include the normalized mean-squared error (NMSE) as a quantitative measure of the difference.

$$NMSE\left(C^{\text{target}}, C^{\text{pred}}\right) = \frac{\sum_{i=1}^{6}\sum_{j=1}^{6}(C^{\text{target}}_{ij} - C^{\text{pred}}_{ij})^2}{\sum_{i=1}^{6}\sum_{j=1}^{6}(C^{\text{target}}_{ij})^2}$$

where $C^{\text{target}}_{ij}$ and $C^{\text{pred}}_{ij}$ are the targeted and predicted elasticity tensor of the generated mechanical metamaterial, respectively. The difference between the stiffness tensors of the target material and the inversely predicted design are measured by examining the components in the elasticity tensor. The relative density (ρ) represents the material ratio per unit volume. The qualitative and quantitative alignment of the elasticity tensors between the targeted topology and our model's generated inverse solutions confirms our model's effectiveness. It not only establishes the utility of our model as a potent tool in the realm of mechanical metamaterials but also underscores its potential to significantly advance generative design methodologies. This alignment between theoretical aspirations and practical outcomes reaffirms our model's role as an innovative cornerstone in the evolving landscape of material science.

**Creating Functionally Graded Mechanical Materials**

Functionally graded mechanical metamaterials (FGMMs) represent a novel class of materials that exhibit tailored properties through controlled variations in their microstructure. FGMMs have garnered significant attention in recent years due to their ability to achieve unconventional mechanical properties derived from their intricate architectures. Traditional materials, often constrained by uniform properties, fall short in satisfying the diverse demands of various applications. In contrast, FGMMs offer the prospect of tailoring mechanical behavior at a local scale, enabling the creation of materials with exceptional and multifunctional properties. Our comprehensive exploration of the property space and the meticulous characterization of latent design variables within our model facilitate the full expression of FGMMs, accommodating a wide array of geometries, densities, and properties.

Figure 7 (a) displays representative examples of FGMMs, with node counts ranging from 20 to 110. The number of nodes correlates with connectivity and structural density. The $C_{11}$, $C_{22}$, and $C_{33}$ values within each interval, normalized against the elasticity of the base material $E_s$, are presented alongside their corresponding graphs and elasticity surfaces. The results highlight the model's capability to produce a seamless transition in elasticity between two designated metamaterial designs, maintaining connectivity even in the middle of varying densities. To confirm the practical applicability of the achieved control over the elastic modulus in mechanical metamaterials, rigorous experimental validation was conducted. A detailed quantitative analysis comparing the model's prediction with actual experimental results, delineated in the Supporting Information, confirms the reliability and feasibility of the predicted data. This validation underscores the efficacy of the proposed modeling framework, highlighting its potential as a generative tool for fabrication of FGMMs.

Figure 7 (b) further demonstrates the controllability of transition layer thickness by adjusting the number of interpolation steps. Target graphs, selected within the node range of 20 to 30, exhibit our model's complete controllability in adjusting the count of transitional mechanical metamaterials from one to seven. Figure 7 (c) underscores the model's fitness in steering the transitional gradient, aligning meticulously with predefined curves. This experiment, manipulating the interpolation gradient between two target graphs, shows that the transitional gradient of the generated FGMMs adheres to two targeted trajectories: sigmoid and linear paths, within preset interpolation steps. This result underscores the potential of the proposed model as a deep generative model for FGMMs. We believe that the ability to create materials with functionally graded mechanical properties of the proposed model offers a paradigm shift in material science, opening up new possibilities for designing structures with unprecedented levels of performance and functionality.

# Conclusion

In this study, we present an innovative model capable of producing metamaterials via elementary arithmetic operations within the latent space, representing a notable stride forward in mechanical metamaterial design. This novel model adeptly explores and exploits the complex design space, achieving a harmonious equilibrium between the simplicity and complexity inherent in deep generative models. Our innovative approach not only elevates the predictive accuracy of the designs but also broadens their applicability across various scenarios. The demonstrated efficacy of our model is underscored by its ability to navigate the intricate landscape of metamaterial's configurations, providing designers with a refined understanding of the interplay between structural features and mechanical properties. By generating a novel metamaterial via simple arithmetic operations, our methodology offers a versatile solution applicable to a broad spectrum of metamaterial design challenges. This work not only propels the field of mechanical metamaterials into a new era but also lays down a foundational framework for future explorations at the confluence of materials science, engineering, and computational modeling. Our findings represent a pivotal contribution to the ongoing evolution of material design methodologies, setting new benchmarks for innovation and applicability.

## Methods

**Database generation**

Figure S1 in Supporting Information elucidates the intricacies of the 14 basis unit cells pivotal in the synthesis of a comprehensive dataset of strut-based lattice structures. These strut-based graphs are initially defined within a cubic domain stretching from 0 to 1. Originating with the 14 fundamental unit cells, an array of novel structures can be generated via linear combinations of these cells. Figure 1 (a) presents the nodal positions and connectivity patterns of the basis unit cells and Figure 1 (b) showcases both the basis unit cells and the array of complex graph structures derived from them. The collective suite of 14 strut-based unit cells, each exhibiting cubic symmetry, facilitates the generation of an expansive set of 16,383 structures through linear combinations, thereby constituting the initial dataset. Based on the initial dataset, new connectivity may be introduced by removing or inserting nodes and edges during the diffusion process as well as latent vector perturbations.

**Generative model framework**

Central to our methodology are deep generative models, leveraging the latent representations they afford for computational efficiency. We incorporate variational autoencoder with discrete latent space, as previously developed in our prior research[41]—a strategy that efficiently mitigates the computational intensity typically associated with latent diffusion models. This efficiency is further enhanced as our diffusion model capitalizes on the discrete latent representation. Moreover, by perturbing the discrete latent vectors, our framework exhibits remarkable extrapolation capabilities, details of which are elaborated in the Supporting Information. Here, we present an overview of our generative framework.

Building on the traditional VAE, the vector quantized (VQ)-VAE introduces a pivotal modification—adopting a discrete latent space. This is achieved through vector quantization, where the encoder's output $z_e(x)$ is mapped to the closest vector in a predetermined codebook, producing a quantized latent representation $z_1(x)$. The process is mathematically captured as $z_q(x) = argmin_i||z_e(x) - e_i||$, where $e_i$ signifies the $i$-th codebook vector. The reconstruction loss, expressed as $\log p(x|z_q(x))$, evaluates the model's capability to reconstruct the input data from the quantized latent vector. Furthermore, a regularization component is incorporated to promote the learning of this discrete representation, enhancing the utilization and diversity of the codebook vectors. On the other hands, score-based diffusion models, particularly Denoising Diffusion Probabilistic Models (DDPM) and Denoising Diffusion Implicit Models (DDIM), further elaborate the generative narrative through a stochastic progression to and from noise distributions[55] . The discrete diffusion process, indexed by a temporal variable $t$, adheres to a SDE. The reverse process, guided by a trained neural network leveraging learned scores, elucidates the denoising trajectory. This robust framework facilitates the methodical conversion of noise into structured, high-fidelity data, highlighting its significance in the generative modeling domain. Building upon the robust foundation of the VQ-VAE, our approach integrates a latent diffusion model that operates synergistically with the Graph Transformer to refine the generative process for graph structures. Contrary to the noise application in image-based diffusion models that occurs on a per-pixel basis, our model adeptly diffuses noise across the distinct nodal and edge features of graphs, categorized by types $\mathcal{X}$ and $\mathcal{E}$. This nuanced diffusion is managed by transition probabilities encapsulated within matrices $[Q_\mathcal{X}^t]_{ij}$ and $[Q_\mathcal{E}^t]_{ij}$, enabling the progressive introduction of noise to the graph's architecture $G^t = (X^t, E^t)$. The Graph Transformer, parameterized by $\theta$, is then employed to reverse this diffusion process, denoising the graph to recover the pristine structure $G$. It achieves this through a sophisticated computation of Queries (Q), Keys (K), and attention weights, which collectively enhance the model's learning capabilities for graph-based tasks. The further details of the proposed algorithm as well as theoretical backgrounds are demonstrated in Supporting Information.

**Inverse Design Methodology**

The initiation of our inverse design process hinges on the specification of a target mechanical property, setting the stage for the application of the latent diffusion process with guidance. In this regard, a vector randomly chosen from the latent space is subjected to a denoising process. This iterative procedure is characterized by the use of an adjusted predictor, integrating the gradient of guidance, a strategy crucial for accurately converging towards the intended solution. This guidance process is described in Algorithm 2 provided in the supplementary information.

We delineate two methodologies for inverse design: the DDIM guidance approach and a perturbation method based on the VQ codebook. The DDIM approach, while stable in generating target graph structures, falls short in extrapolation performance compared to the perturbation method. Conversely, the perturbation method, though superior in extrapolation, risks degrading graph connectivity, potentially resulting in ill-conditioned graph structures. To harness the strengths of both, we employ a combined strategy for optimal outcomes. For DDIM guidance, conditioning on a target value $C_0$ involves introducing a joint score function. More precisely, given a conditional target value $C_0$, the joint score function can be defined as

$$\nabla_{v_t} \log\left(p_\theta(z_t) p_\phi(C_0|z_t)\right) = \nabla_{z_t} \log p_\theta(z_t) + \nabla_{z_t} \log p_\phi(C_0|z_t) = -\frac{\epsilon_\theta(z_t)}{\sqrt{1-\alpha_t}} + \nabla_{z_t} \log p_\phi(C_0|z_t),$$

where $\alpha_t$ is a time-dependent coefficient that modulates the balance between the initial data and the noise in the latent space. It starts near 1 at $t = 0$, maintaining the structure of the original graph. As $t$ increase to 99, $\alpha_t$ decreases toward 0, allowing the noise to dominate, effectively randomizing the latent vector and making the conditional probability $p(z_t|z_0)$ converge to a standard normal distribution. $p_\theta$ is the score function network, $p_\phi$ the regressor network, and $C_0$ corresponds to a selected material property from $\{C_{11}, C_{22}, C_{33}\}$. This function is then integrated into the DDIM iterations, replacing the standard score function to facilitate conditional generation aligned with the specified material property.

To detail the VQ perturbation method, we initiate by selecting $J$ vectors, denoted as $n_j$, from a standard normal distribution $\mathcal{N}(0, \mathrm{I})$ for $1 \leq j \leq J$. These vectors are transformed into codebook vectors and passed through the decoder network to generate graph structures. The material properties, determined as ground truth by a numerical solver, guide the subsequent selection of a latent vector $\hat{n}_j$ that closely matches the targeted property $C_0$, yielding $\hat{C}$. The process iterates with perturbations of $\hat{n}_j$ by $\epsilon_2 \mathcal{N}(0, \mathrm{I})$ in the latent space until the difference $|\hat{C} - C_0|$ is less than a defined threshold $\epsilon_1$. This iterative process is thoroughly outlined in Algorithm 1, provided in the Supporting Information.

**Creating Functionally Graded Mechanical Metamaterials**

Our approach utilizes a systematic strategy to interpolate between graph pairs from the database, focusing on graphs with similar node densities. Node density intervals range from 10-20 to 100-100 nodes. Within each interval, two graphs, termed Graph A and Graph B, are selected based on their comparable node densities. Graph A and B undergo encoding to derive discrete latent vectors (pre-Vector Quantization (VQ), denoted as $z_0$ in the training process). These vectors are denoted as Latent A and Latent B. These vectors are subjected to a Euclidean norm computation to ensure suitability for interpolation. Pairs with norms under 20 are chosen, mitigating potential complications in property interpolation for higher norms. Utilizing the interpolation formula derived from the reference[56], the latent vectors between Latent A and B can be obtained. Subsequent to interpolation, the latent vector undergoes processing via Vector Quantization (VQ) and a Decoder. This step is crucial for constructing the graph's architecture, detailing node locations and their interconnections. Node information from Graphs A and B is strategically combined, existing on the proximity of the intermediate latent vector to each. Illustratively, in a 10-step interpolation sequence, node data from Graph A is incorporated for the initial five steps, transitioning to Graph B's data from the sixth step onwards. This process adheres to the Node Attribute framework delineated in the training process. The interpolation's efficacy is further evaluated through the computation of $C_{11}$, $C_{22}$, and $C_{33}$ values for the resultant graph, employing a Matlab numerical solver. The selection criterion for the optimal interpolation scenario is the demonstration of the most linear progression in these values. Such a meticulous approach is pivotal for ensuring the precision and effectiveness of interpolations among targeted graph structures in the database. The main purpose of this sophisticated procedure is the creation of metamaterial designs that are not only mechanically robust but also have unique and inventive features.

**Data Availability**

The data that support the findings of this study are available from the corresponding authors on reasonable request.

**Acknowledgment**

The work of Y. Hong was supported by Basic Science Research Program through the National Research Foundation of Korea (NRF) funded by the Ministry of Education (NRF-2021R1A2C1093579) and by the Korea government (MSIT) (RS-2023-00219980). The work of N. Kim was supported by the National Research Foundation of Korea (NRF) grant funded by the Korean government (MSIT) (No. 2022R1C1C1009387, No. 2022R1A4A3033320). This study was also supported by the National Supercomputing Center with Supercomputing Resources, Republic of Korea, including technical support (KSC-2023-CRE-0303).

**Author Contributions**

**Conflict of Interest**

The authors confirm that there is no conflict of interest to declare.

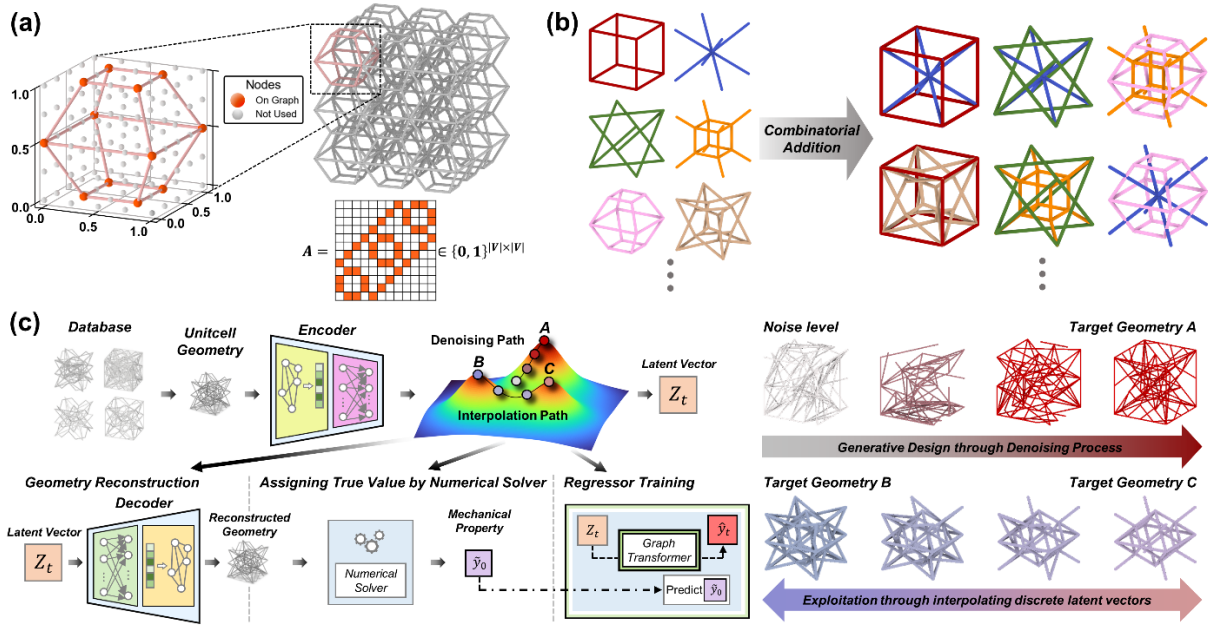

**Figure 1.** An overview of database construction from fundamental unit cells and their subsequent training process over the latent space. (a) The selection of nodes for a basic unit cell from a predefined $5 \times 5 \times 5$ grid, where fixed geometric positions define node candidates, and their interconnections establish the fundamental graph structure. (b) This depicts how each unit cell in the database emerges from a combinatorial synthesis of 14 basic unit cells, yielding a total database size of $2^{14} - 1 = 16{,}383$, excluding the null graph. (c) The continuous transformation is displayed along the noising/denoising trajectory within the latent space. In this space, the DDIM model navigates a generative denoising pathway, where a noised latent vector, corresponding to a disordered graph structure, converges to a latent vector representing a structured unit cell geometry.

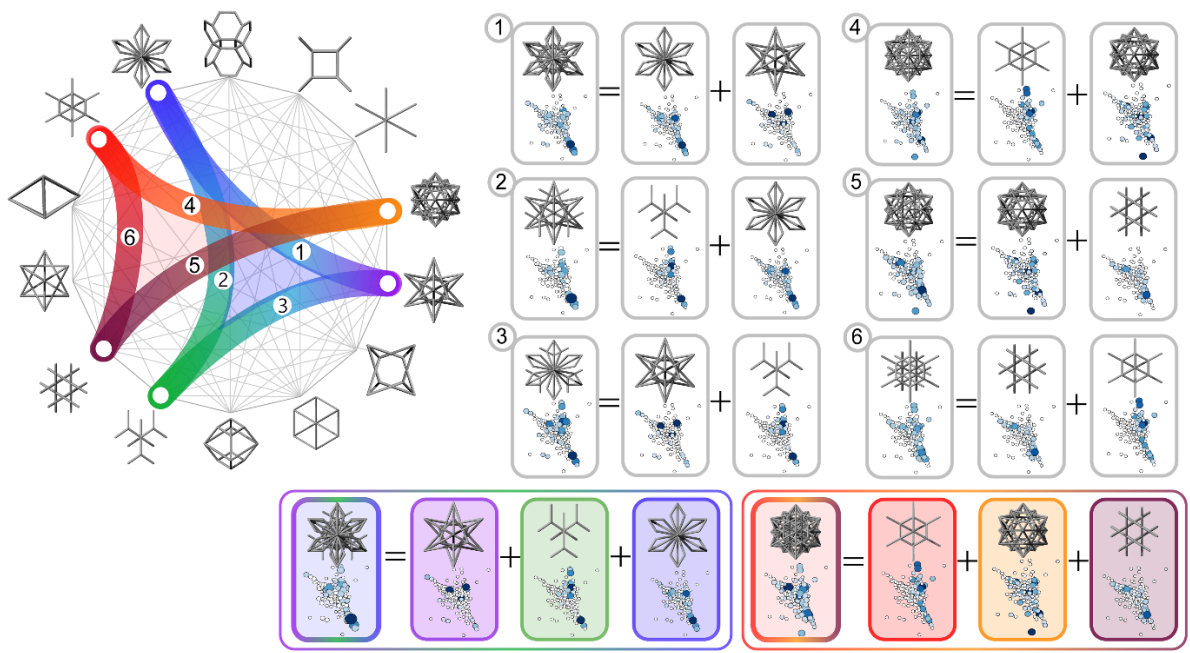

**Figure 2.** Combinatorial synthesis of graph structures from basis latent. The network diagram illustrates 14 basis graphs. Adjacent plots present individual and combined codebooks. Lines connecting the basis graphs denote the combinations of basis latent that yield the graph overlapping one, with the triangle indicating specific instances of two or three basis latent combination.

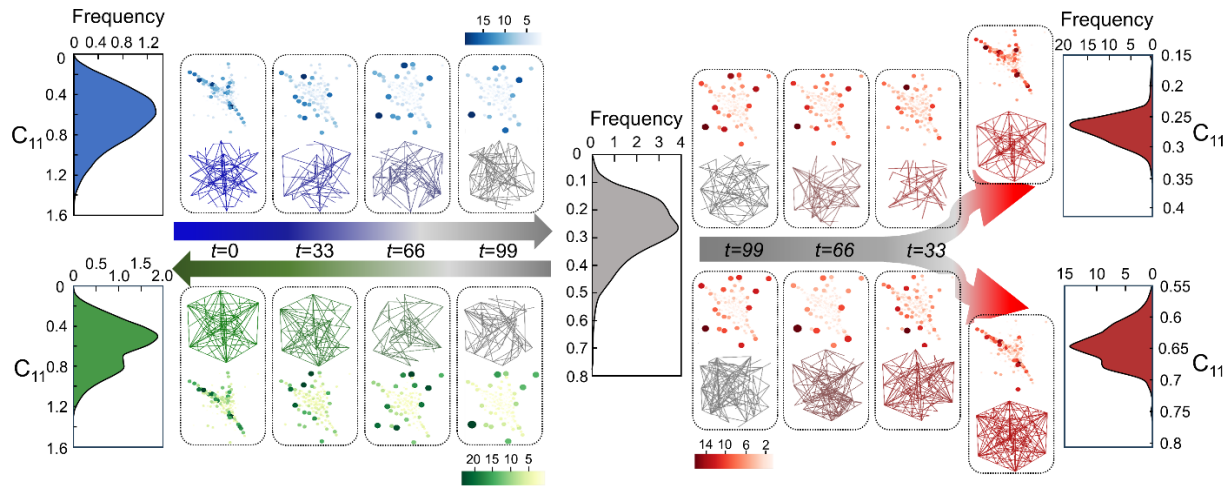

**Figure 3.** This diagram visualizes the dynamic evolution of the distribution of mechanical property $C_{11}$ alongside representative graph structures throughout the diffusion process. The blue distribution exemplifies the initial values of $C_{11}$ drawn from our database. As the forward process unfolds, depicted by the grey distribution at $t = 99$, $C_{11}$ values spread due to noise infusion across 500 graph samples. Conversely, the green distribution captures the reverse denoising trajectory, commencing from a standard normal distribution $\mathcal{N}(0, \mathrm{I})$ and converging towards noise-free graph structures. The guided red distributions reflect the outcomes of targeted denoising processes, which steer the $C_{11}$ values towards specified targets of 0.28 and 0.66, demonstrating the effectiveness of guidance.

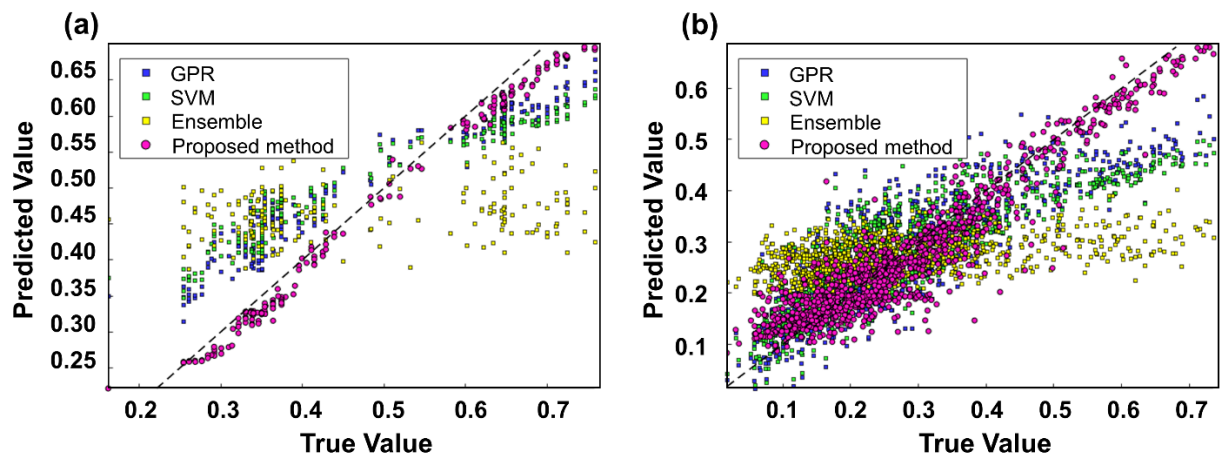

**Figure 4.** Comparison of performance in predicting material properties $C_{11}$. (a) Noise-free graphs at $t = 0$ and (b) noised graphs over the range $0 < t < 100$.

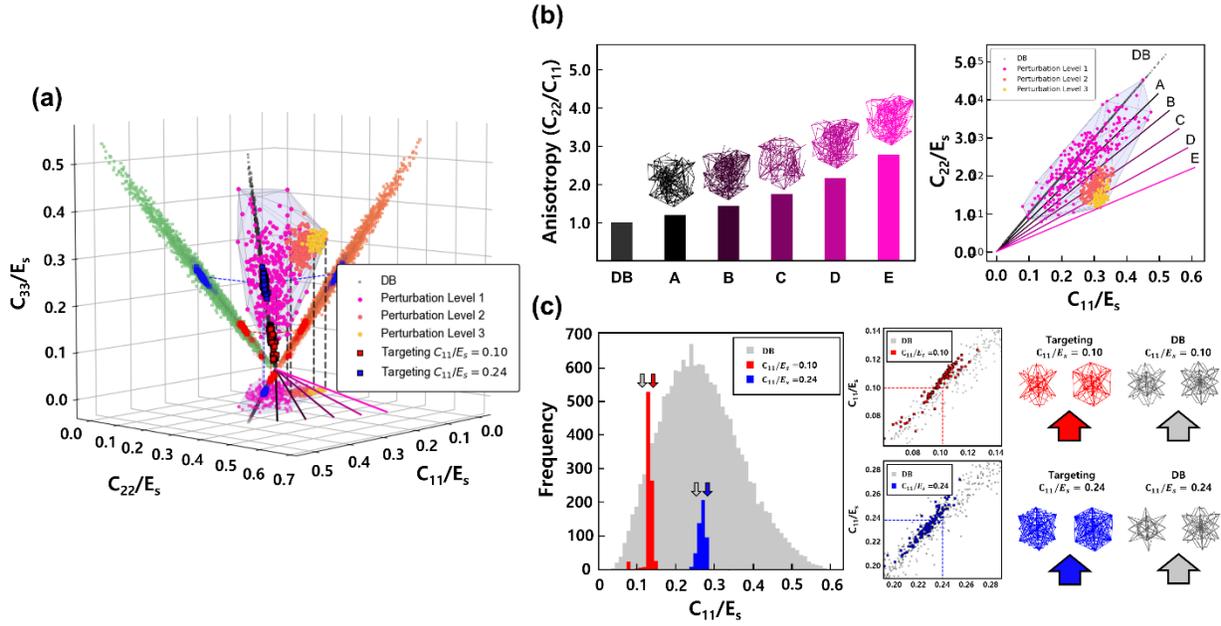

**Figure 5.** Exploration and exploitation on the mechanical property space. (a). The construction of the augmented property space is demonstrated where exploration via VQ perturbation (denoted by pink and yellow circles) and exploitation using DDIM with guidance (indicated by red and blue squares) enrich the diversity of mechanical properties. (b). The VQ perturbation process is presented to probe the extremes of topological configurations with Poisson's ratios extending from 1 to 2.75, capturing values beyond the existing database. (c) This highlights the augmentation of property space, where DDIM with guidance is harnessed to prolifically generate unit cell structures targeting normalized stiffness parameters of $\bar{C}_{11} = 0.1$ or $\bar{C}_{11} = 0.24$.

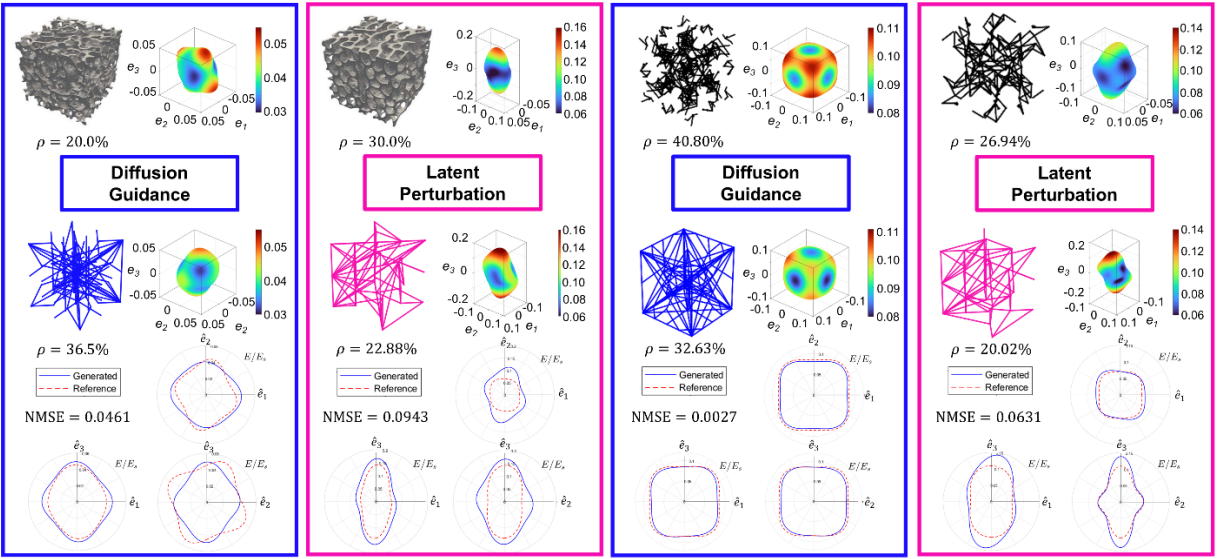

**Figure 6.** This diagram displays a selection of inverse design candidates and their elastic property visualizations generated through DDIM with guidance and latent space perturbation methods. The first set illustrates trabecular bone topologies inspired by Collabella et al. (2017), while the latter pair presents designs based on the Crystallographic Periodic Networks studied by Lumpe and Stankovic (2021). The upper section of the figure depicts the target geometries alongside their mechanical property profiles. Central to the figure are the inverse design candidates, strategically situated to bridge the theoretical and resultant topologies. Below, we present the elastic modulus surfaces projected onto Euclidean coordinate axes, offering a comparative view of the design objectives against the generated structures' performance.

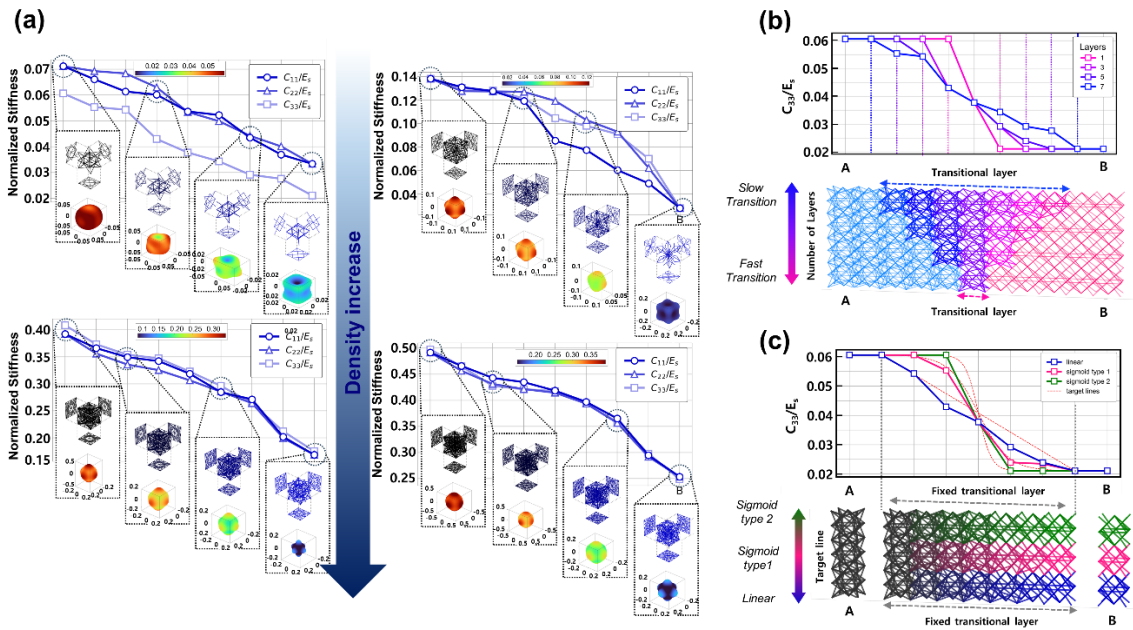

**Figure 7.** Morphological transitions design for functionally graded material. (a) A gradational shift in unit cell geometries correlated with the escalation of node density is displayed alongside the consequent mechanical property variations. (b) This captures the dynamic interplay between topological transformations and the normalized stiffness across various transitional layers. (c) The curve fitting of mechanical property is demonstrated across a spectrum of geometric transitions within a fixed transitional layer.

# Supporting Information

**Simple arithmetic operation in latent space can generate a novel three dimensional graph metamaterials**

*Namjung Kim[1], Dongseok Lee[2], Chanyoung Kim[2], Dosung Lee[1] and *Youngjoon Hong[2]

[1]Department of Mechanical Engineering, Gachon University, Sungnam, 13120, Korea

[2]Department of Mathematics, Korea Advanced Institute of Science and Technology, Deajeon, 34141, Korea

*Corresponding author
Prof. Youngjoon Hong:    hongyj@kaist.ac.kr
Prof. Namjung Kim:    namjungk@gachon.ac.kr

**Geometry of 14 basis unit cells corresponding graph representations**

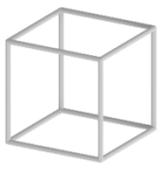

**Figure S1.** Fourteen basis unit cells were used to construct the initial database.

**Experimental validation through compression test**

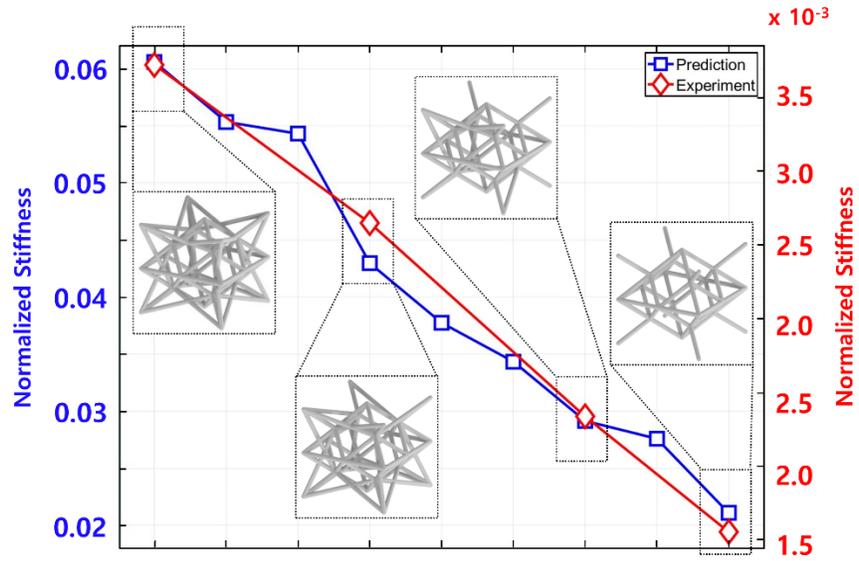

**Figure S2.** Comparison between the predicted stiffness values and the experimental results from the compression test. Four graph structures with 26 nodes are selected as the specimens. The Y-axis is $C_{11}$ values normalized using Young's modulus of the base material. Due to the initial fluctuations in the experimental setup, Young's modulus is measured at the strain ε=4.2% after the transient state ends.

**Stress-strain curves of four representative samples in compression test**

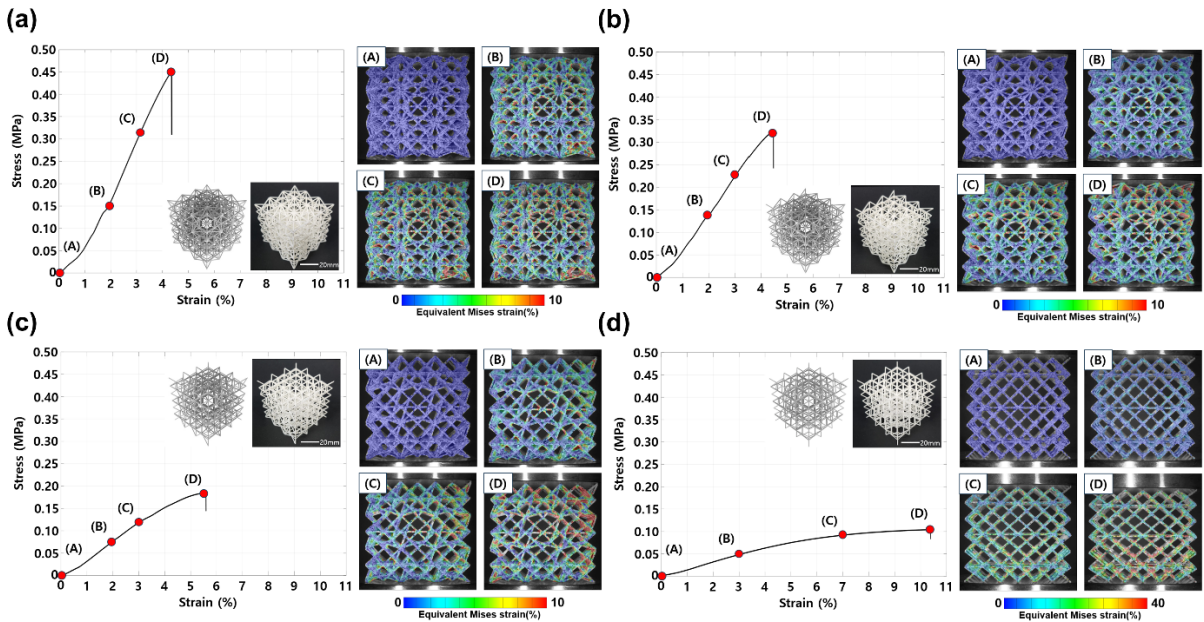

**Figure S3.** Compression test stress-strain curves for four specimens are presented, highlighting the feasibility and reliability of the proposed modeling framework. A commercially available 3D printer, the Formlab 3, with UV-curable resin used to fabricate the samples. The fabrication parameters were rigorously defined to ensure the lattice structures were accurately manufactured. A 2x2x2 supercell was created to obtain the elastic modulus of the sample in the experimentally viable size. The compression tests were meticulously conducted using a universal testing machine (Technology & Dolf Ltd) equipped with a 10 kgf load cell, applying a small strain step to prevent dynamic effects on the specimens. The deformation patterns were captured and analyzed using the digital image correlation (DIC) method, providing insightful visualizations of the deformation responses.

**Sequential training process for the proposed BVBM framework**

The proposed deep generative model framework is trained in four subsequent steps. Figure S4 (a) provides a comprehensive illustration of the initial training process for the vector-quantized variational autoencoder (VQ-VAE) with a discrete latent representation augmented by the complex structural capabilities of graph neural networks (GNNs). Our model is adept at processing input data consisting of graphs synthesized intricately from a combination of fourteen basis unit cells. Initially, the encoder acts as the primary gateway for input data. The encoder is designed to integrate a standard graph convolutional network (GCN) followed by a fully connected (FC) network, which further compresses these aggregated features into a dense latent vector, effectively reducing the dimensionality while retaining the essential characteristics of the input data. After the encoding phase, the latent space is discretized. This pivotal step leverages a predefined codebook comprising numerous discrete vectors. The latent vector is quantized by mapping it to the nearest vector within this codebook, thereby transforming the continuous properties of the latent space into a discrete form denoted as $Z_t$. In the final stage, the decoder reconstructs geometric data from a quantized latent representation. The architecture of the decoder mirrors that of the encoder, but operates in reverse, progressively reconstructing the data. A notable difference from conventional models is that our decoder does not prioritize the prediction of node positions within the graph. Rather, it is tailored to discern the edge connectivity between nodes, a strategy made possible by the fact that the node positions are already established and remain invariant within the latent space. The efficacy of our model was underscored by its capacity to encode and decode intricate graph data, thereby extending the capabilities of conventional generative models. This facilitates advanced applications in machine learning, which necessitate discerning the underlying connectivity patterns within the data.

Figure S4 (b) depicts the training process employed in the latent diffusion model. The previously trained encoder was used to provide a latent representation $Z_0$ of the given mechanical metamaterial design. Within the latent space, the model implements a forward process that transforms the initial latent representation $Z_0$, into the evolved state $Z_t$. This transformation is called the diffusion process within the latent space and converts $Z_0$ into a noisy state, which is represented by $Z_t$. The subsequent critical step in the training stage is the graph Transformer, which is a specialized neural network designed to process graph-structure data. Using the noisy latent state $Z_t$ as the input, the key objective of the graph transformer is to reverse the diffusion process, which is called the denoising phase of diffusion. The comprehensive nature of the training, demarcated by the dotted outlines within the diagram, indicates a rigorous learning phase in which the latent diffusion model internalizes the complex interplay between noisy and denoised states in the latent space. This capability is crucial for generating new data samples that faithfully replicate the statistical properties of the original dataset. For metamaterial generation, this model leverages the latent diffusion process to exploit the design space of the materials. This step aims to use a distorted latent representation to accurately regenerate the initial data configuration. Upon acquiring $Z_0$, the subsequent stages, including the decoder network, aligned with the procedures delineated in the previously mentioned VAE with a discrete latent space.

Figure S4 (c) illustrates the third key training process, which is identified as regressor training. This step is crucial for defining the guidance conditions for our diffusion model. The process begins with an encoder, in which the unit cell geometry serves as the primary input. This graph input is adeptly managed by the encoder, comprising a GCN followed by an FC, condensing the extracted features into a compact, lower-dimensional latent space representation, aptly denoted by $Z_0$. When transitioning into the latent space, the model undergoes a forward diffusion process that dynamically transforms $Z_0$ into $Z_t$, representing a modified state of the latent representation as it progresses through time or sequential processing steps. Within a clearly defined regressor training module, a graph Transformer is tasked with the critical function of predicting $y_t$, which is a mechanical property derived from the latent representation of the given design $Z_t$. This predictive task is fundamental for the training and calibrating the guidance mechanism of the diffusion model. In parallel, the decoder reconstructs the design of the mechanical metamaterial from the latent representation $Z_t$, mirroring the reverse encoder steps. The reconstructed geometry of the mechanical metamaterial was then directed into a numerical solver, which is an analytical tool designed for numerical computations. This solver aims to numerically predict the mechanical properties of a given metamaterial, $y_t$, thus validating and reinforcing the predictions made by the machine learning model with numerical solutions.

**(a)**

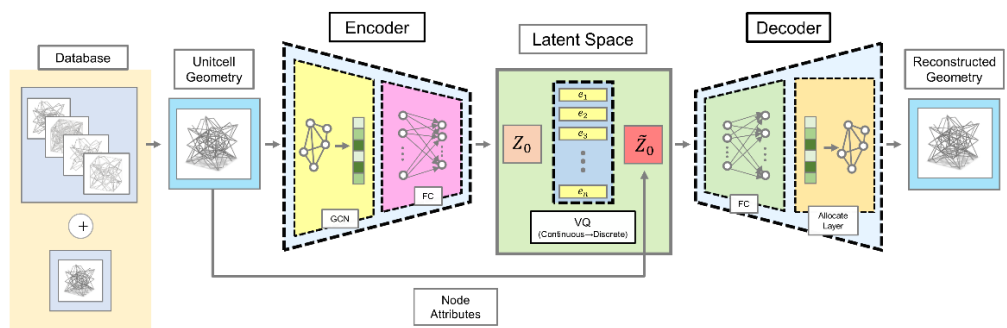

**(b)**

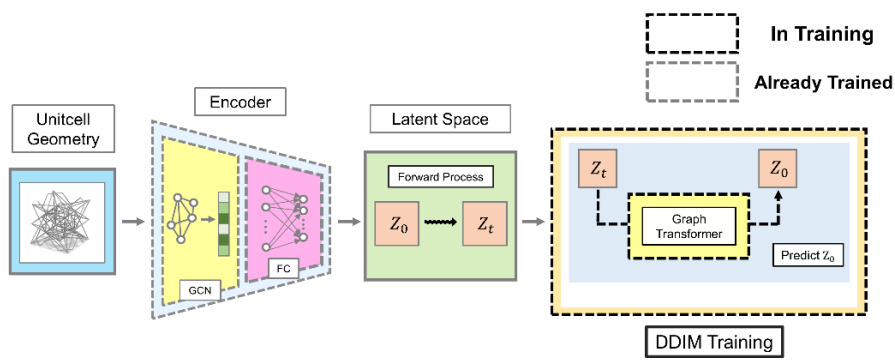

**(c)**

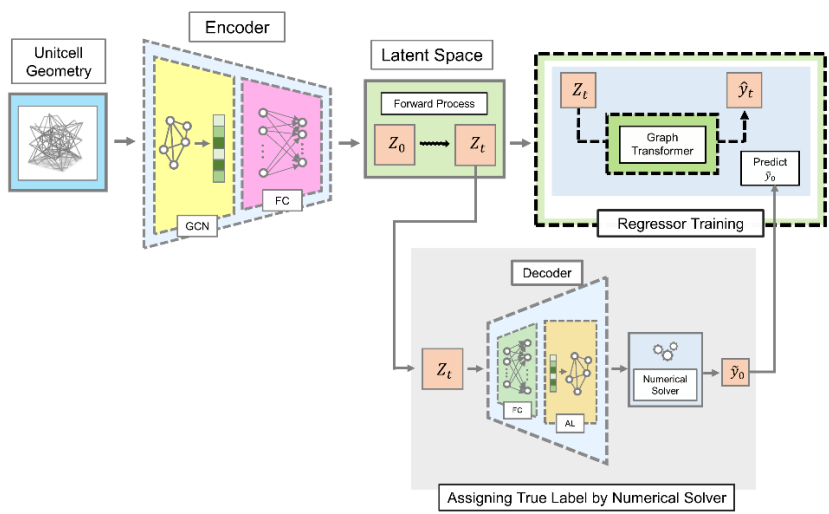

**Figure S4.** Schematic diagram of training process. (a) Latent space design for encoder, decoder and VQ training. (b) DDIM training process over the trained latent space. (c) Regressor training process for guidance generation.

**Conditional generation process with guidance**

Figure S5 shows a schematic representation of the sampling framework, which incorporates a latent diffusion process along with a discrete codebook representation. The process begins with the input of the unit cell geometry, presumably referring to the fundamental geometric structures that form the basis of the model. These inputs were processed, as shown in Figure S4 (b). Subsequently, the model progressed to the latent space, where the $Z_0$ representation underwent a diffusion process, as indicated by the forward arrow, leading to a noisy latent state $Z_t$. The Graph Transformer, marked in green within the latent space, is crucial. This is based on the following equation:

$$p_\phi(y|Z_t) = \frac{e^{-(C(y-reg_\phi(Z_t))^2)}}{K},$$

The Graph Transformer utilizes the noised state $Z_t$ to predict $Z_{t-1}$, where $reg_\phi(Z_t)$ is obtained from the regression model shown in Figure S4 (c). Here, $K$ is a normalizing constant, which makes integral of $p_\phi$ equal to one by defining

$$K = \int_{\mathbb{R}^{N \times 18}} e^{-(C(y-reg_\phi(Z_t))^2)} dZ_t,$$

and the correction factor $C \in \mathbb{R}$ determines how sharply the probability density $p_\phi(y|Z_t)$ peaks at the regression value $y \in \mathbb{R}$. This mechanism is essential for the denoising phase and follows the conditions imposed by the variable $y$, which are derived from the learned data distribution of the model. In the latent space, trained vector quantization was applied to obtain a realigned denoised latent representation. After completing the framework, the decoder attempts to reconstruct the original geometry from the denoised latent representation. The model successfully generated a detailed and high-fidelity geometric representation from the learned data distributions, showcasing its ability to produce new samples.

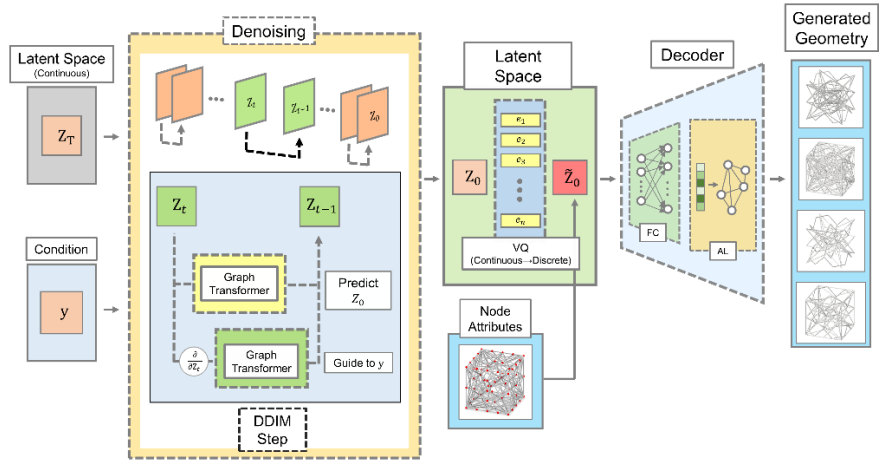

**Figure S5.** Conditional generation process with guidance

**Inverse design for various lattice structures**

Figure S6 demonstrates the application of our method in the inverse design of various unit cell structures. The structures on the top of Figure S6 include from the crystallographic periodic network (Lumpe and Stankovic, 2021), alongside two unitcells derived from the reference (Mohammad Abu-Mualla and Jida Huang, 2023). The corresponding elastic surfaces are also shown on the right side of the geometry in the figure. The graph metamaterials inferred by DDIM alongside their relative density ρ, with the corresponding elastic surfaces and projections of Young's modulus illustrating that the DDIM model adeptly navigates the inverse design space to identify geometries with mechanical properties akin to the targeted benchmarks. The similarity of the mechanical properties between the target geometries and topologies inferred by DDIM was verified by the projection plots of the elastic surface at the bottom of the figure. The projection plots included three planes, which are $e_1 - e_2$, $e_1 - e_3$, and $e_2 - e_3$. The efficacy of the DDIM model was quantitatively confirmed using the computed normalized mean square error (NMSE) for the proposed inverse solutions.

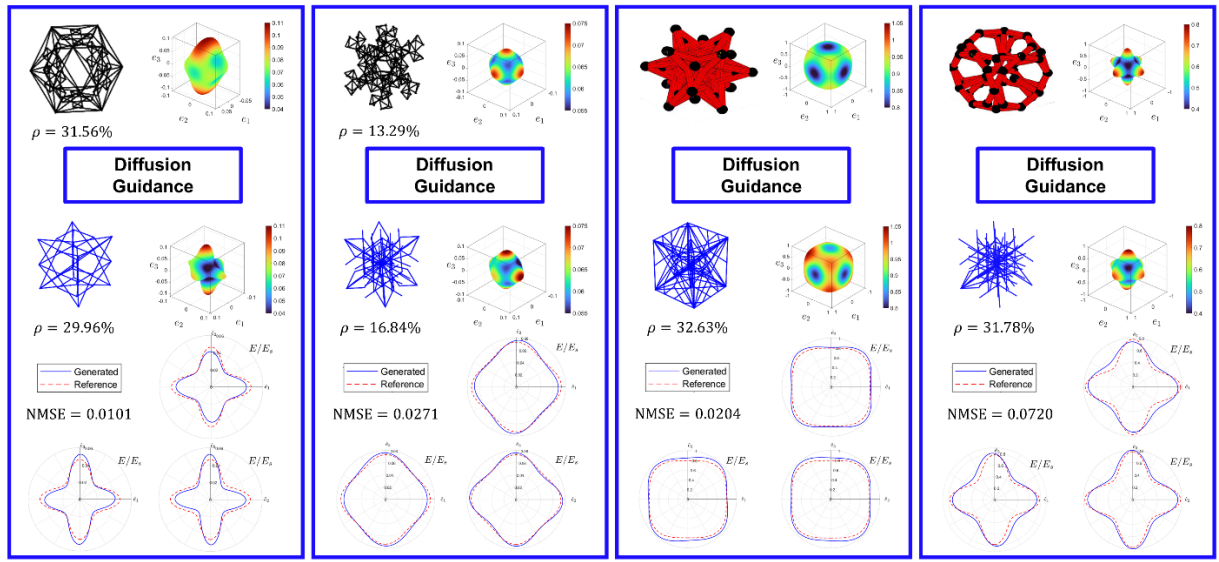

**Figure S6.** Inverse design for various unitcell structures. Figure depicts the representative geometries, along with their corresponding elastic surfaces and projections of Young's modulus, derived from inverse design solutions.

**Description of VQ-VAE and diffusion model (DDIM)**

**VAE and VQ-VAE.** The Variational Autoencoder (VAE) framework is designed to establish a stochastic relationship between observed data space, denoted as $x$, and a latent space, $z$. This learning process is interpreted as a directed model characterized by the joint distribution $p_\theta(x, z) = p_\theta(x|z) p_\theta(z)$, where $\theta$ represents the model's parameters, and $p_\theta(z)$ is the prior distribution over latent variables. Although the conditioned distribution $p_\theta(x|z)$ is parameterized by a decoder, it is generally not tractable. To address this, the VAE employs an encoder, another deep neural network, to map x to z by approximating the posterior distribution. The training's likelihood function becomes tractable through the evidence lower bound (ELBO) defined as:

$$\text{Loss} = \mathbb{E}_{q(z|x)}[\log p(x|z)] - \mathbb{D}_{KL}[q(z|x)||p(z)],$$

where $\mathbb{D}_{KL}$ stands for the Kullback-Leibler divergence. The loss function of the VAE thus comprises two key components: the reconstruction loss $\mathbb{E}_{q(z|x)}[\log p(x|z)]$ assessing the model's reconstruction accuracy, and the regularization loss $\mathbb{D}_{KL}[q(z|x)||p(z)]$, which ensures conformity of the learned latent representation to the prior distribution. On the other hand, VQ-VAE introduces label information into the latent space, leading to a deterministic and constrained representation of the learned data. The VQ-VAE, in particular, differs from the VAE in its approach to latent representation. While a VAE learns a continuous latent representation, a VQ-VAE learns a discrete one. This discrete representation is achieved through a process known as vector quantization. The discrete latent embedding space, denoted as $e \in R^{N \times D}$, can be considered as a set of codebooks $\{e_1, e_2, \ldots e_N\}$, where each $e_i$ is a vector in $R^D$. During the encoding phase, an input $x$ is first mapped to a latent representation $z_e(x)$ by the encoder network. The VQ-VAE then assigns $z_e(x)$ to the nearest vector in a predefined set, resulting in a quantized latent representation $z_q(x)$. This process is mathematically expressed as

$$z_q(x) = argmin_i ||z_e(x) - e_i||,$$

where $e_i$ denotes the i-th vector in the codebook. The reconstruction loss in VQ-VAE, given by

$$\log p\left(x|z_q(x)\right),$$

measures how well the model can reconstruct the input data from this quantized latent representation. Furthermore, the regularization term in VQ-VAE, written as

$$||sg[\,z_e(x)] - e||_2^2 + \beta||z_e(x) - sg[e]||_2^2,$$

plays a crucial role in learning the discrete representation. Here, [·] stands for the stop gradient operator. The term $||z_e(x) - sg[e]||_2^2$ encourages the encoder's output to be close to one of the codebook vectors, while $||sg[\,z_e(x)] - e||_2^2$ helps maintain the diversity of the codebook vectors. Overall, VQ-VAE's discrete latent space allows for more structured and interpretable representations compared to the continuous latent space of standard VAEs, making it well-suited for tasks with inherent discrete structures or categories.

**Diffusion model.** Score-based diffusion models, such as Denoising Diffusion Probabilistic Models (DDPM) and Denoising Diffusion Implicit Models (DDIM), are based on forward and reverse Stochastic Differential Equations (SDEs). These models effectively transform data into and from a noise distribution through a stochastic process. We define a discrete diffusion process consisting of random variables indexed by time variable $t$ in the interval $[0,1]$, denoted as $\{z_t\}_{t=0}^T$, where $T$ represents the total number of time steps. This process is assumed to follow a given SDE, under the condition that the functions $f$ and $g$ exhibit sufficient regularity. The SDE, which describes the forward evolution of the process in time, is expressed as:

$$dz = f(z,t)dt + g(t)dw,$$

where $w$ represents a standard Wiener process, also commonly referred to as Brownian motion. This equation encapsulates the dynamic nature of the diffusion process, driven by the drift function $f$ and the diffusion coefficient $g$, with the element of randomness introduced by the Wiener process. The forward SDE describe the gradual addition of noise to the data. Initially, data at $z_0 \sim p_0$ is noised according to the SDE, and the conditional probability density $p(z_t, t|z(0), 0)$ following the SDE is governed by the Fokker-Planck equation:

$$\frac{\partial p(z_t, t|z_0, 0)}{\partial t} = -\nabla_z^T[f(z,t)p(z_t, t|z_0, 0)] + \frac{1}{2}\nabla_z^T[gg^T p(z_t, t|z_0, 0)]\nabla_z,$$

where $\nabla_z^T$ is the differential operator $\left[\frac{\partial}{\partial z^{(1)}}, \ldots, \frac{\partial}{\partial z^{(N)}}\right]$. Assuming $p(z_t, t)$ reaches a steady state at time $T$, $p(z_T, T)$ is defined as the prior distribution. Starting from the prior distribution $p(z_T, T)$ and reverting to $p(z_0, 0)$, the reverse SDE is defined as

$$dz = [f(z,t) - g(t)^2 \nabla_z \log p_t(z)]dt + g(t)dw.$$

This reverse process demonstrates that we can sample data similar to the original data distribution $p(z_0, 0)$ from the prior distribution $p(z_T, T)$, as indicated by the solution to the diffusion process. DDPM and DDIM leverage these principles to generate high-quality, diverse samples, forming a crucial part of the broader class of generative models that utilize diffusion and denoising in stochastic processes. We consider the DDPM through the lens of a Markov chain $\{z_0, z_1 \ldots, z_T\}$ where each transition $p(z_i|z_{i-1})$ is modeled as a Gaussian distribution $\mathcal{N}(\sqrt{1-\beta_i}z_i, \beta_i I)$ with $0 < \beta_i < 1$. It is possible to express the conditional probability as

$$p(z_t|z_0) = (\sqrt{\bar{\alpha}_t}z_0, (1-\bar{\alpha}_t)I),$$

where $\bar{\alpha}_t = \prod_{i=1}^{t} \alpha_i$. This indicates that as data becomes increasingly noised, it converges to the prior distribution $p(z_T|z_0) \sim \mathcal{N}(0, I)$. From the discrete Markov chain formulation

$$z_i = \sqrt{1-\beta_i}z_{i-1} + \sqrt{\beta_i}z_\mathcal{N}$$

with $z_\mathcal{N} \sim \mathcal{N}(0, I)$ for $i = 1, 2 \ldots T$, we derive the forward SDE as

$$dz = -\frac{1}{2}\beta(t)zdt + \sqrt{\beta(t)}dw,$$

where $\beta\left(\frac{i}{T}\right) = T\beta_i$. The reverse time SDE is then formulated as

$$dz = \left(-\frac{1}{2}\beta(t)z - \beta(t)\nabla_z \log p_t(z)\right)dt + \sqrt{\beta(t)}dw.$$

As the exact marginal probability $p(z_t, t)$ is typically unknown, the score function $\nabla_z \log p_t(z)$ is approximated by considering the reverse process with $p(z_{t-1}|z_t, z_0) = \mathcal{N}(\mu_t(z_t, z_0), \sigma_t)$, where

$$\mu_t(z_t, z_0) = \frac{1}{\sqrt{1-\beta_t}}(z_t + \beta_t \nabla_z \log p_{t|0}(z_t|z_0)).$$

By defining a neural network-based distribution $p_\theta(z_{t-1}|z_t) = \mathcal{N}(\mu_{\theta,t}(z_t), \sigma_t)$, the mean of the normal distribution $\mu_{\theta,t}(z_t)$ is defined by

$$\mu_{\theta,t}(z_t) = \frac{1}{\sqrt{1-\beta_t}}(z_t + \beta_t s_\theta(z_t)).$$

Then, we minimize the loss function

$$L(\theta) = \sum_{t=1}^{T} \gamma_t \mathbb{E}_{z(0)} \left[ \left\| \epsilon - \epsilon_\theta \left( \sqrt{\bar{\alpha}_t} z_0 + \sqrt{1 - \bar{\alpha}_t} \epsilon, t \right) \right\|^2 \right],$$

where $\epsilon_\theta(z_t, t) = -\sqrt{1 - \bar{\alpha}_t} s_\theta(z_t)$, $\bar{\alpha}_t = \prod_{s=1}^{t}(1 - \beta_s)$, $\epsilon \sim \mathcal{N}(0, \mathrm{I})$ and $\gamma_t$ are weighting coefficients. Upon training the score based neural network $s_\theta(z_t)$, we can implement the reverse discrete sampling process by

$$z_t = \left(1 + \frac{\beta_{t+1}}{2}\right) z_{t+1} + \beta_{t+1} s_\theta(z_{t+1}) + \sqrt{\beta_{t+1}} z_{t+1,\mathcal{N}},$$

where $z_{t,\mathcal{N}} \sim \mathcal{N}(0, \mathrm{I})$ and $1 \leq t \leq T - 1$, enabling the generation of data samples similar to the original data distribution $p(z_0, 0)$ from the prior distribution $p(z_T, T)$. In contrast to the DDPM, the DDIM introduces a unique inference distribution defined by the variable $\sigma_t \in R$. The distribution $q_\sigma(z_{1:T}|z_0)$ in DDIM is formulated as:

$$q_\sigma(z_{1:T}|z_0) = q_\sigma(z_T|z_0) \prod_{t=2}^{T} q_\sigma(z_{t-1}|z_t, z_0).$$

For the convenience we use notation $\prod_{s=1}^{t}(1 - \beta_s)$ as $\alpha_t$, where each conditional distribution $q_\sigma(z_{t-1}|z_t, z_0)$ is a Gaussian distribution defined as

$$q_\sigma(z_{t-1}|z_t, z_0) = \mathcal{N}\left(\sqrt{\alpha_{t-1}} z_0 + \sqrt{1 - \alpha_{t-1} - \sigma_t^2} \frac{(z_t - \sqrt{\alpha_t} z_0)}{\sqrt{1 - \alpha_t}}, \sigma_t^2 \mathrm{I}\right).$$

From this, we can derive the forward transition $q_\sigma(z_t|z_{t-1}, z_0)$ as:

$$q_\sigma(z_t|z_{t-1}, z_0) = \frac{q_\sigma(z_{t-1}|z_t, z_0) q_\sigma(z_t|z_0)}{q_\sigma(z_{t-1}|z_0)}.$$

Joint probabilistic density fuction for generative process in DDIM is defined as $p_\theta(z_{0:T}) = p_\theta(z_T) \prod_{t=1}^{T} p_\theta(z_{t-1}|z_t)$, where:

$$p_\theta(z_{t-1}|z_t) = \begin{cases} \mathcal{N}(f_\theta(z_1), \sigma_1^2 \mathrm{I}) \; if \; t = 1 \\ q_\sigma(z_{t-1}|z_t, f_\theta(z_t)) \; otherwise \end{cases},$$

and $f_\theta(z_t)$ is given by:

$$f_\theta(z_t) = \frac{z_t - \sqrt{1 - \alpha_t} \epsilon_\theta(z_t)}{\sqrt{\alpha_t}}.$$

Especially we can generate a sample $z_0$ starting from $z_T \sim \mathcal{N}(0, I)$, iteratively in the following way

$$z_{t-1} = \sqrt{\alpha_{t-1}} f_\theta(z_t) + \sqrt{1 - \alpha_{t-1}} \epsilon_\theta(z_t),$$

which is deterministic sampling because there is no random noise term. Even though the diffusion process is characterized as non-Markovian, the loss fuction optimized in DDIM framework remain identical to that used in DDPM. The loss fuction is given by

$$L(\theta) = \sum_{t=1}^{T} \tilde{\gamma}_t \mathbb{E}_{z_0} \left[ |\epsilon - \epsilon_\theta(\sqrt{\bar{\alpha}_t} z_0 + \sqrt{1 - \bar{\alpha}_t} \epsilon, t)|^2 \right].$$

**Description of VQ perturbation sampling and DDIM with guidance sampling**

We have a sampling algorithms using the latent space containing a wider range of graphs than DB space and is capable of generating new graphs to get desired material property $C$ value, where $C \in \{C_{11}, C_{22}, C_{33}\}$. In particular, Algorithm 1 is composed of VQ-perturbation that enables extrapolation, and there are two distinct sampling processes with their own advantages: Algorithm 1 for generating extrapolated graphs with the desired $C$ value and Algorithm 2, the DDIM+guidance algorithm, which also enables the generation of graphs with the desired $C$ value.

---

**Algorithm 1** Sampling from VQ-perturbation
---
**Input** : $n_j \sim N(0, I),\ 1 \leq j \leq J$
**Targeting** $C_0$
$\hat{n} \leftarrow vq(argmin_{n_j}\{|C_0 - C(dec(vq(n_1)))|, \ldots |C_0 - C(dec(vq(n_J)))|\})$
$\hat{C} = C(dec(vq(\hat{n})))$

**while** $|C_0 - \hat{C}| > \epsilon_1$ **do**
   Sample $n_j \sim \hat{n} + \epsilon_2 N(0, I),\ 1 \leq j \leq J$
   $\hat{n} \leftarrow vq(argmin_{n_j}\{|C_0 - C(dec(vq(n_1)))|, \ldots |C_0 - C(dec(vq(n_J)))|\})$
   $\hat{C} = C(dec(vq(\hat{n})))$
**end for**
**Output** $\hat{n}$

---

To actively utilize this space, we introduce Algorithm 1, which iteratively perturbs to generate graphs with the desired $C$ value. First, we generate J samples $n_j$ from $\mathcal{N}(0, I)$, and then set the target material property value $C_0$. For each sample, we calculate $|C_0 - C(dec(vq(n_j)))|$. Here, $vq(\cdot)$ denotes the vector quantized operator to the codebook, and $dec(\cdot)$ denotes the decoder operator that transforms into a graph. The material properties solved by a numeric solver for the input graph $G$ is denoted as $C(G)$. With a small threshold value $\epsilon_1 \sim 0.1$, if the material properties value of the current $\hat{n}$ is not sufficiently close to $C_0$, we perturb $\hat{n}$ by small noise term $\epsilon_2 \mathcal{N}(0, I)$ with $\epsilon_2 \sim 0.1$ to gradually move close to $C_0$.

**Algorithm 2** Sampling from DDIM+guidance

**Input:** $z_T \sim N(0, I)$
**for** $t = T$ **to** 1 **do**
$$\hat{\epsilon} \leftarrow \epsilon_\theta(z_t) - \sqrt{1-\alpha_t}\nabla_{z_t} \log p_\phi(C_0|z_t)$$
$$z_{t-1} \leftarrow \sqrt{\alpha_{t-1}}\left(\frac{z_t - \sqrt{1-\alpha_t}\hat{\epsilon}}{\sqrt{\alpha_t}}\right) + \sqrt{1-\alpha_{t-1}}\hat{\epsilon}$$
**end for**
**Output** $z_0$

---

Algorithm 2 applies a sequence of transformations to a graph sample initially sample from $\mathcal{N}(0, I)$, using a joint score fuction to progressively shape it into a graph that matches the desired target value $C_0$. Especially the joint score function is defined as

$$\nabla_{z_t} \log\left(p_\theta(z_t)p_\phi(C_0|z_t)\right) = \nabla_{z_t} \log p_\theta(z_t) + \nabla_{z_t} \log p_\phi(C_0|z_t) = -\frac{\epsilon_\theta(z_t)}{\sqrt{1-\alpha_t}} + \nabla_{z_t} \log p_\phi(C_0|z_t).$$

In our approach, we define the probabilistic model for guidance sampling using the equation

$$p_\phi(y|z_t) = \frac{e^{-C(y-reg_\phi(z_t))^2}}{K},$$

where $C$, $K$ are introduced at explanation of Figure S5 and $reg_\phi(\cdot)$ is graph transformer regressor that takes $z_t$ as input to predict material properties $y$.

**Regressor based on Graph Transformer**

Similar to how diffusion models add noise to each image pixel independently, our method introduces noise to each graph node and edge feature individually. Given the expansive state space of graphs, which precludes the construction of a full transition matrix, we focus on node types $\mathcal{X}$ and edge types $\mathcal{E}$. The transition probabilities for a node and respectively an edge, are encapsulated by the matrices $[Q_\mathcal{X}^t]_{ij} = q(x_j^t | x_i^{t-1})$ and $[Q_\mathcal{E}^t]_{ij} = q(e_j^t | e_i^{t-1})$. Therefore, the diffusion process involves the addition of noise to the graph structure $G^t = (X^t, E^t)$ by sampling from a categorical distribution for each node and edge type. The denoising neural network, parameterized by $\theta$, takes a noisy graph $G^t = (X^t, E^t)$ as input and is trained to infer the clean graph $G$. This is a crucial step for effectively reversing the diffusion process. The architecture of the network, inspired by the transformer model, computes queries (Q) and keys (K) and derives attention weights by transforming the input graph features $X$. Attention is scaled and normalized, focusing on the most relevant information that enhances the learning capability of the network for graph-based tasks. Our graph transformer is designed to handle the intricacies of graph structures by leveraging self-attention mechanisms to compute attention scores and contextually weighted representations. It is characterized by row-wise permutation invariance, which ensures that the model remains robust to graph rotations and permutations of the latent vector, thus preserving the predictive accuracy for noise.

**Comparison study between VQ-VAE and diffusion models**

In our study, a comparative analysis of the VQ-VAEs and diffusion models revealed the distinct advantages inherent to each method. Through experimental validation, we discerned that the diffusion models exhibit certain strengths over VQ-VAE in specific contexts, and vice versa. A critical aspect of this comparison is the engineering of connected 3D unit graphs, particularly when assembling multiple graph cells along the x,y,and z axes. The occurrence of unconnected nodes at interfaces poses significant challenges, potentially undermining the structural integrity of the assembled unit. To evaluate a graph's connectivity systematically, we introduced a metric to assess the connectivity of a graph, denoted as G, which is defined as follows:

$$Connectivity(G) == \frac{2}{3N}\sum_{i=1}^{N}\sum_{j\in\{x,y,z\}}\frac{2}{3N}\sum_{j\in\{x,y,z\}}\frac{|P_j^+ \cap P_j^-|}{|P_j^-|},$$

where $P_j^+$ represents the subset of nodes within graph $G$ characterized by their $j$-th coordinate having a value of one and $P_j^-$ denotes the subset where the $j$-th coordinate is -1. This metric serves as a quantifiable measure, enabling us to ascertain the robustness of the internode connections within the graph and ensure the seamless integration of multiple graph cells in a 3D assembly.

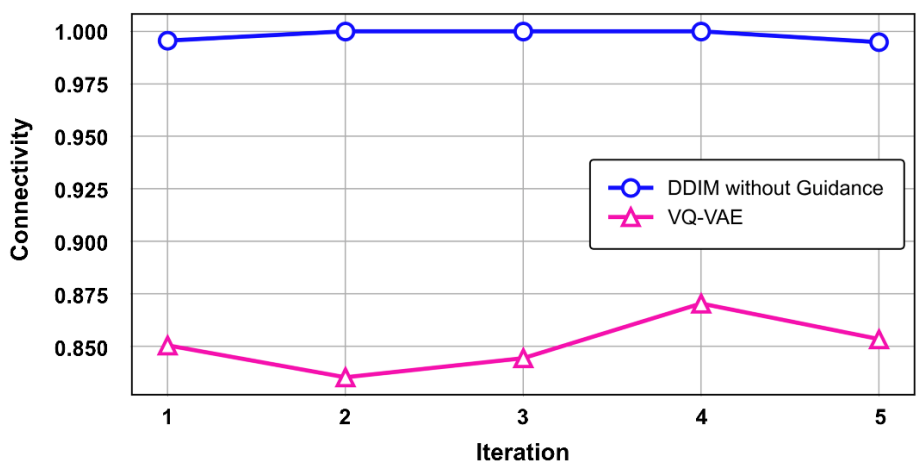

**Figure S8.** Comparison in connectivity of DDIM without guidance and VQ-VAE for generation.

Figure S7 illustrates the results of the comparative analysis of connectivity in randomly generated samples using each algorithm. We employed a batch-based approach in which 100 samples were generated by both DDIM and VQ-VAE in each iteration. Connectivity served as the primary assessment metric. The results, as depicted in the figure, underscore the superior performance of DDIM, which consistently maintains a perfect connectivity score of one throughout all iterations. In contrast, samples generated by VQ-VAE displayed variability in their connectivity scores, typically hovering at approximately 0.85. This variance highlights the relative robustness of DDIM in producing structurally coherent samples across successive iterations.

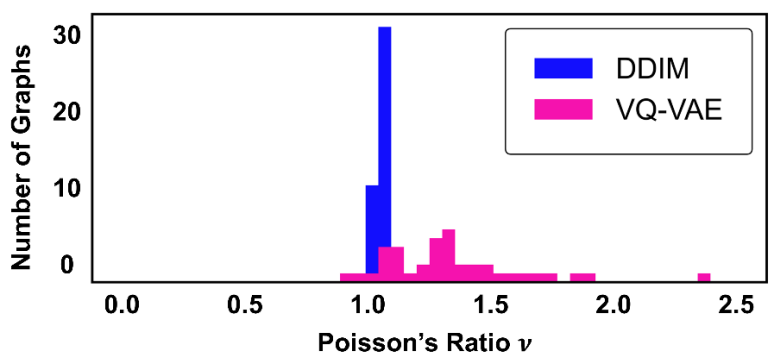

**Figure S9.** Distribution of Poisson's ratio $v$ of the unitcells generated by DDIM and VQ-VAE.

Figure S8 illustrates the distribution of the Poisson's ratio values for a set of 40 graphs, each exhibiting a connectivity value of one generated by the DDIM and VQ-VAE algorithms. The Poisson's ratio, denoted as $v$ and defined by $v = C_{12}/C_{22}$ serves as a critical indicator of diversity within the generated graph structures. The histogram presented in Figure S9 reveals a distinct pattern: while DDIM predominantly produces graphs with $v$ values clustering around one, indicating a hallmark of similarity to the training dataset, VQ-VAE demonstrates a wider range in Poisson's ratio. This broader distribution indicates the capability of the VQ-VAE to generate a more diverse array of graph structures. The potential of VQ-VAE to extrapolate beyond the training dataset and generate novel graph structures was highlighted, showcasing its ability to maintain structural connectivity while introducing variance and breaking symmetry in the generated graphs.

**Geometrical evolution in noising and denoising processes**

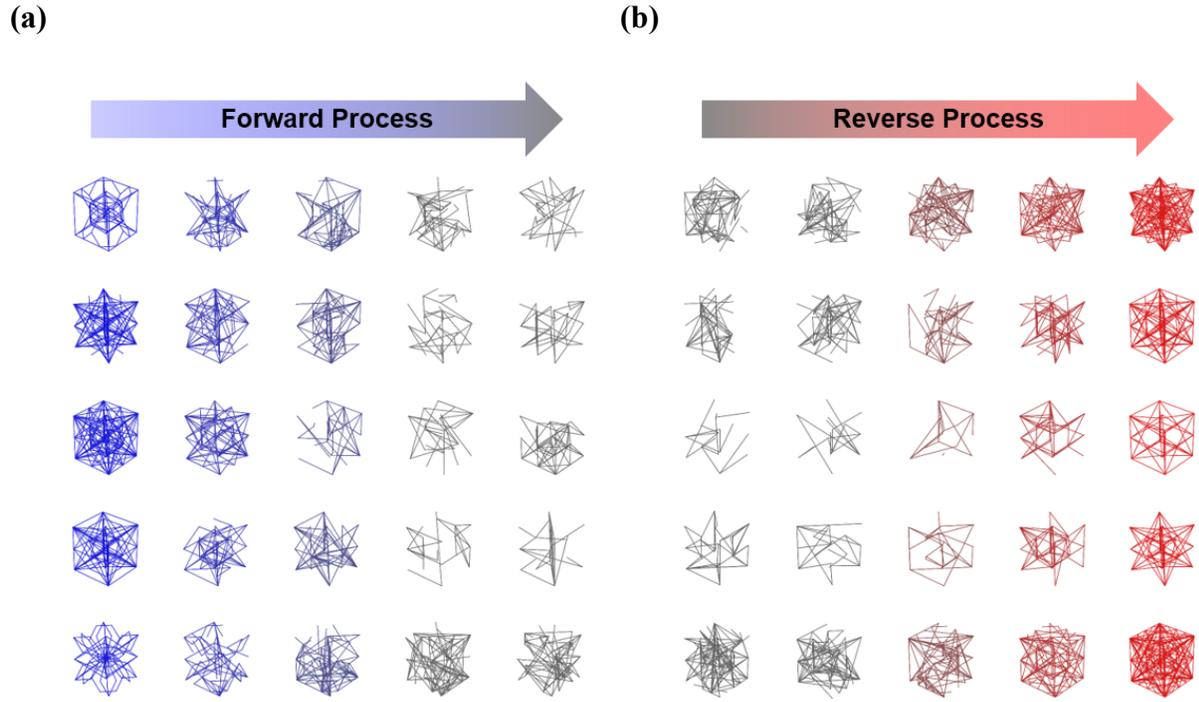

**Figure S10.** Graph evolution of forward process and reverse process. (a) A sequence of graph samples $G_t$ that are noised according to the forward process $p(z_t | z_0) = \mathcal{N}(\sqrt{\alpha_t} z_0, (1 - \alpha_t)I)$. (b) The samples are subjected to a reverse denoising process, wherein the sequence progressively reverts to a noise-free graph state, achieved through the iterative implementation of the specified equation $z_{t-1} = \sqrt{\alpha_{t-1}} f_\theta(z_t) + \sqrt{1 - \alpha_{t-1}} \epsilon_\theta(z_t)$, starting from $z_T \sim \mathcal{N}(0, I)$.

**Latent space interpolation**

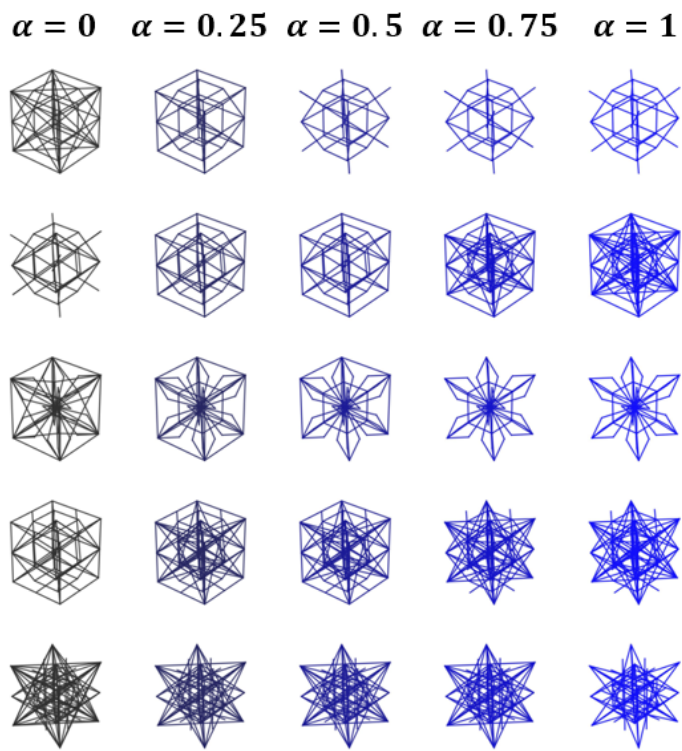

**Figure S11.** Latent space interpolation. The figure shows a sequence of graphs after performing interpolation on $\mathcal{N}(0, \mathrm{I})$ in the following manner $z_T^\alpha = \frac{\sin((1-\alpha)\theta)}{\sin\theta} z_T^0 + \frac{\sin(\alpha\theta)}{\sin\theta} z_T^1$, where $\theta = arccos\left(\frac{(z_T^0)^T z_T^1}{\|z_T^0\|\|z_T^1\|}\right)$, and subsequently it undergoes a denoising process. This procedure involves interpolating between two graphs positioned in the leftmost and rightmost columns.

**Unit cell geometries with the relative elastic modulus 0.1**

| Geometry | 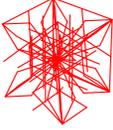 | 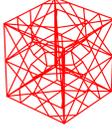 | 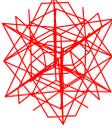 | 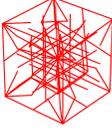 |
|---|---|---|---|---|
| **Targeting** $\bar{C}_{11} = 0.10$ | 0.0987 | 0.1031 | 0.1159 | 0.0829 |
| **Connectivity** | 1.00 | 1.00 | 1.00 | 0.861 |

**Figure S12.** Examples of DDIM guidance generation targeting $\bar{C}_{11} = 0.1$. Various graph generated by using DDIM with guidance targeting material property $C_{11} = 0.28$, which is the same as targeting $\bar{C}_{11} = 0.10$, defined as $\bar{C}_{11} = \frac{C_{11}}{E_s}$ and $E_s$=0.28. Even when including asymmetric graphs, the generated structures exhibit a consistently high level of connectivity.